\documentstyle[aps]{revtex}
\begin{document}
\draft
\title{Coherent pairing states for the Hubbard model}
\author{A. I. Solomon\footnote[1]{Permanent address: Faculty of Mathematics and Computing, 
Open University, Milton Keynes, MK7 6AA, U. K.   email:a.i.solomon@open.ac.uk}}
\address{Laboratoire de Gravitation et Cosmologie Relativistes,\\
 Universit\'{e} Pierre et Marie Curie - CNRS, \\
URA 769, Tour 22, 4-\`{e}me \'{e}tage, Bo\^{\i}te 142,\\
4, place Jussieu, 75252 Paris Cedex}
\author{ K. A. Penson\footnote[2]{email: penson@lptl.jussieu.fr}}
\address{Laboratoire de Physique Th\'{e}orique des Liquides, \\
Universit\'{e} Pierre et Marie Curie - CNRS, \\
URA 765, Tour 16, 5-\`{e}me \'{e}tage, Bo\^{\i}te 121,\\
4, place Jussieu, 75252 Paris Cedex}
\maketitle
\begin{abstract}
We consider the Hubbard model and its extensions on bipartite lattices. We
define a dynamical group based on the $\eta$-pairing operators introduced
by C. N. Yang, and define coherent pairing states, which are combinations of
eigenfunctions of $\eta$-operators. These states permit exact calculations
of numerous physical properties of the system, including energy, various
fluctuations and correlation functions, including pairing ODLRO to all
orders. This approach is complementary to BCS, in that these are
superconducting coherent states associated with the exact model, although
they are not eigenstates of the Hamiltonian.
\end{abstract}
\pacs{PACS numbers:74.20.-z,05.30.Fk, 71.27.+a}
The Hubbard model plays a special role in condensed matter physics. It
allows one, in appropriate limits, to model the electronic properties of
systems ranging from insulators to superconductors. It is generally believed
that  high-$T_c$ superconductivity may be described by some form of the
Hubbard model. Although the model cannot be solved except in one dimension,
some insight into its properties in general dimensions can be obtained
through the so called $\eta$-pairing mechanism introduced by C. N. Yang 
\cite{Yang89}. This mechanism  allows one to construct a subset of
the exact spectrum of the model. The eigenfunctions obtained through $\eta$-pairing possess the 
property of off-diagonal long-range order (ODLRO)
and thus are superconducting.
In this note we introduce a new family of wave functions which are
combinations of $\eta$-pairing eigenfunctions. The  $\eta$-pairing procedure has been applied to 
a number of strongly-correlated fermion systems\cite{Essler}-\cite{Sato}. Our wavefunctions
are coherent pairing states (CPS) of the dynamical group of the Hubbard
model. Although not eigenfunctions of the Hamiltonian, they permit exact
calculations of numerous physical properties of the Hubbard model, including
the energy, arbitrary moments of the Hamiltonian, fluctuations and
correlation functions, including ODLRO which is shown to be non-vanishing. The CPS are 
mathematically related to the variational wave functions
used in a mean field treatment of  BCS type\cite{BCS}.\\ \\
{\underline{$\eta$-pairing}} For the Hubbard model we adopt the definition and notation of 
Yang\cite{Yang89}. Let $a_{ \vec{r}}^{+}$
and 
$b_{\vec{r}}^{+}$
be real-space creation operators for spin-up and spin-down electrons
respectively, i.e. $c_{\vec{r}\uparrow }^{+}$= $a_{
\vec{r}}^{+\text{ }}$, $c_{\vec{r}\downarrow }^{+}$= $
b_{\vec{r}}^{+\text{ }}$ with $a_{\vec{r}}^{+\text{ }}$
and $b_{\vec{r}}^{+\text{ }}$ satisfying the usual fermion
anti-commutation relations.

Consider a 3D Hubbard model on a $L\times L\times L=M$ cube (L even) with
periodic boundary conditions. The Hamiltonian is given by
\begin{eqnarray}
H&=&T_0+T_1+V
\\
T_0&=&A\epsilon \sum_{\vec{k}}(a_{\vec{k}}^{+}a_{
\vec{k}}^{}+b_{\vec{k}}^{+}b_{\vec{k}}^{}) \label{t0}
\\
T_1&=&-B\sum_{\vec{k}}(\cos k_x+\cos k_y+\cos k_z)(a_{
\vec{k}}^{+}a_{\vec{k}}^{}+b_{\vec{k}
}^{+}b_{\vec{k}}^{}) \label{t1}
\\
V&=&2W\sum_{\vec{r}}a_{\vec{r}}^{+}a_{\vec{r}
}^{}b_{\vec{r}}^{+}b_{\vec{r}}^{},
\end{eqnarray}
where $\epsilon >0$, $a_{\vec{k}}^{+\text{ }}$ is the Fourier
transform of $a_{\vec{r}}^{+\text{ }}$ , $2W$ is the on-site
Hubbard interaction of arbitrary sign, and $A$ and $B$ arbitrary constants. We
introduce the $\eta$-operators which create (annihilate) a fermion pair
with momentum $\vec{\pi }$:
\begin{equation}
\eta =\sum_{\vec{r}}e^{i\vec{\pi }\vec{r}
}a_{\vec{r}}b_{\vec{r}}=\sum_{\vec{k}}a_{
\vec{k}}b_{\vec{\pi }-\vec{k}},
\end{equation}
\begin{equation}
\eta ^{+}=\sum_{\vec{r}}e^{i\vec{\pi }\vec{r
}}a_{\vec{r}}^{+}b_{\vec{r}}^{+}=\sum_{\vec{
k}}b_{\vec{\pi }-\vec{k}}^{+}a_{\vec{k}}^{+}.
\end{equation}
It has been shown (\cite{Yang89}, \cite{Nowak81}) that the operator $\eta ^{+}$ satisfies 
\begin{equation}
\left[ H,\eta ^{+}\right] =E\eta ^{+},  \label{spec}
\end{equation}
with $E=2A\epsilon +2W$. Equation ( \ref{spec}) is typical of a Spectrum Generating Algebra 
(\cite{Dothan},\cite{Goshen})   and implies that  for any power-expandable 
$f(\eta^{+})$
\begin{equation}
\left[ H,f(\eta ^{+})\right] =E\eta ^{+}f^{\prime }(\eta ^{+}).
\label{specfn}
\end{equation}
Note that E does not depend on B. The relation (\ref{spec}) for the Hubbard
model was derived some time ago\cite{Nowak81} but its consequences were only fully exploited  
by Yang\cite{Yang89}. The operators $\eta$
satisfy the angular momentum commutation relations of SU(2):
\begin{equation}
\left[ \eta ^{+},\eta \right] =2\eta _z  \label{SU2}
\end{equation}
\begin{eqnarray}
\eta _z &=&\frac 12 \sum_{\vec{r}}(n_{\vec{r}}^{(a)}+n_{
\vec{r}}^{(b)}-1)  \nonumber \\
&=&\frac 12\sum_{\vec{r}}n_{\vec{r}} -\frac{1}{2}M,
\end{eqnarray}
where the local occupation number $n_{\vec{r}}$ is equal to $n_{
\vec{r}}^{(a)}+n_{\vec{r}}^{(b)}=a_{\vec{r}
}^{+}a_{\vec{r}}^{}+b_{\vec{r}}^{+}b_{\vec{r
}}^{}.$ 
We also observe that from Eq.(\ref{SU2}) that the following relation holds
\begin{equation}
\left[ \frac{\eta ^{}}{\sqrt{M}},\frac{\eta ^{+}}{\sqrt{M}}\right] =1-d,
\label{SU2mod}
\end{equation}
where $d $ is the electronic density, $d =M^{-1}(\sum_{\vec{
r}}n_{\vec{r}})$. Equation (\ref{SU2mod}) indicates that for
small electron density the operators $\frac \eta {\sqrt{M}}$ are
approximately bosons\cite{Lipkin73}. The operators $\eta $ also satisfy the
relations $\left( \eta ^{+}\right) ^{M+1}=\left( \eta \right) ^{M+1}=0$,
reflecting, according to the  Pauli-principle, the impossibility of occupying a given site $\vec{r\text{ 
}}$ by more than one pair $(a^{+}b^{+})$. For given
$M$ and using (7) one can produce $M$ exact, normalized eigenstates of H by
applying succesive powers of $\eta ^{+}$ on the vacuum state $|vac\rangle $.
So
\begin{equation}
|\Psi _N\rangle =\beta (N,M)(\eta ^{+})^N|vac\rangle ,\; \; \; \; N=1,...,M
\end{equation}
is a simultaneous eigenstate of H and of the operator N$_2$ counting the
number of doubly occupied sites, N$_2=\sum_{\vec{r}}n_{
\vec{r}}^{(a)}\cdot n_{\vec{r}}^{(b)}$, 
\begin{eqnarray}
H|\Psi _N\rangle &=&NE|\Psi _N\rangle  \nonumber \\
N_2|\Psi _N\rangle &=&N|\Psi _N\rangle ,  \label{N2}
\end{eqnarray}
where $\beta (N,M)$ is a normalization factor equal to\cite{Yang89} 
\begin{equation}
\beta (N,M)=\left[ \frac{ (M-N)!}{M! N!}\right] ^{\frac 12}.
\end{equation}
Evidently, $\langle \Psi _N|(\eta ^{+})^r|\Psi _N\rangle =\delta _{r,0}$.
Note that $\left[ H,N_2\right] \neq 0$. We observe that the $
|\Psi _N\rangle $ depend neither  on the value nor on the  sign of W. In general, $\left[\eta , 
H\right] \neq0$ except for the half-filled band\cite{Zhang}. \\ \\
{\underline {Dynamical Group for H}  }
  With this in mind, we embed the Hamiltonian H together with $\vec{\eta }
=\left\{ \eta ,\eta ^{+},\eta _z\right\} $, in a larger, dynamical,  group. Define a new operator $J_0$ 
by 
\begin{equation}
J_0=\frac HE-\eta _z.
\label{J0}
\end{equation}
Using (\ref{spec}) and its hermitian conjugate we find that 
\begin{equation}
\left[ J_0,\vec{\eta }\right] =0.  \label{J0eta}
\end{equation}
We conclude that the smallest group containing  H is $\left\{ J_0,
\vec{\eta }\right\} $, where $J_0$ is the center of the group but
\underline{not}  the unit operator. The dynamical group 
of our Hubbard model is thus U(2). This would appear to be the first instance of a dynamical 
group for an {\em exact} interacting many-body system. The relation (\ref{J0eta}) is
essential for the calculation of any expectation values of H. \\ \\
{\underline{Coherent Pairing States}  }
We introduce a normalized spin  coherent state by 
\begin{eqnarray}
|\mu \rangle &=&{\cal N}^{-\frac 12}{e}^{ \mu \eta} | 0 \rangle  \nonumber  \\
	      &=&(1 + |\mu |^2)^{-\frac{M}{2}} {e}^{ \mu \eta} | 0 \rangle
\label{nscs}
\end{eqnarray}
where the state $|0 \rangle $ is the filled pair state $|0 \rangle = {\frac{1}{M!}}({\eta}^{\dagger})^M 
|{\rm vac}\rangle $.  We refer to $|\mu>$ as a {\em coherent pairing state}. This step is 
reminiscent of the BCS wave function\cite{BCS}, which is however not related to any Hamiltonian 
with a local potential energy.  In contrast, our states arise out of the exact relations 
Eqn.(\ref{spec}) and Eqn.(\ref{J0eta}).
  In the limit $M\rightarrow \infty $, $ {\vert}\mu \rangle $
becomes an eigenstate of $\frac \eta {\sqrt{M}}\;$and apart from
normalization is a harmonic oscillator coherent state \cite{Radcliffe}.
The state $|\mu \rangle $ is not an eigenstate of H. In contrast with 
$| \Psi _N\rangle$ it involves components with different
numbers of particles (pairs) and thus gives rise to  
non-zero values of $\langle \mu |\eta ^r|\mu \rangle$. Further, using 
$\;$(\ref{spec}), (\ref{specfn}) and (\ref{J0eta}) we may calculate 
$\langle \mu |H^p|\mu \rangle $ for any $p=1,2,3,...$
in terms of  $\langle \mu |\eta ^r|\mu \rangle $.
We first calculate $\langle \mu |H|\mu \rangle $ by purely
algebraic means:
\begin{eqnarray*}
H|\mu \rangle &=&{\cal N}^{-\frac 12}\left\{\left[ H,e^{\mu \eta
}\right] +e^{\mu \eta }H\right\} {\frac{1}{M!}}(\eta ^{+})^M|vac\rangle \\
&=&{\cal N}^{-\frac 12}\left\{-E\eta \mu e^{\mu \eta }+e^{\mu \eta}ME\right\} {\frac{1}{M!}}(\eta 
^{+})^M|vac\rangle \\
&=&(-\mu E\eta +ME)|\mu \rangle .
\end{eqnarray*}
where ${\cal N(\mu)}=(1+|\mu |^2)^{M}.$
The required expectation value $\langle \mu |H|\mu \rangle $ becomes $
ME-\mu E\langle \mu |\eta |\mu \rangle $, which, using the results of \cite
{Radcliffe} (Formula 4.2), leads to 
\begin{equation}
\langle \mu |H|\mu \rangle =\frac{ME}{(1+|\mu |^2)}.  \label{muhmu}
\end{equation}
The formula (\ref{muhmu}) indicates that the energy of the state $\vert \mu \rangle $ (which 
involves different numbers of pairs) is equal to
the energy of the fully filled state ${\vert}0\rangle $ (ME)
reduced by the factor $(1+|\mu |^2)^{-1}\leq 1$. The physical
meaning of the parameter $\mu$ is obtained from the 
average number of pairs in a state ${\vert}\mu \rangle$, 
$\langle \mu |N_2|\mu \rangle $. Since ${\vert}\mu \rangle $ does
not depend on the Hamiltonian's parameters, 
\begin{equation}
\langle \mu |N_2|\mu \rangle =\frac 12\langle \mu |\frac{\partial H}{
\partial W}|\mu \rangle =\frac 12\frac M{1+|\mu |^2}\frac{\partial E}{
\partial W}.  \label{n2trick}
\end{equation}
We conclude that 
\begin{equation}
|\mu |^2=\frac {1}{{\overline{n}}_2}-1,  \label{mun2}
\end{equation}
where ${{\overline{n}}_2}=\frac{\langle N_2\rangle }M$ is the average density of pairs in
the state ${\vert}\mu \rangle $.
We may extend the set of states for which exact analysis is available by introducing  $r$-depleted 
states, defined by (normalized) 
\begin{equation}
{\vert} \mu;r \rangle ={{\cal N}_{r}} ^{-\frac 12}\eta ^r   {\vert}\mu \rangle .
\end{equation}
These are analogues of the displaced number states of quantum optics\cite
{Knight}. These states give rise to a more interesting energy spectrum than the equidistant Yang 
case, with the gap between neighbouring depleted states  ${\vert} \mu;r \rangle$ and $ {\vert} 
\mu;r-1 \rangle$ being given by 
\begin{equation}
\triangle _r(|\mu |^2)=\frac{\langle \mu |(\eta ^{+})^{r-1}H\eta ^{r-1}|\mu
\rangle }{\langle \mu |(\eta ^{+})^{r-1}\eta ^{r-1}|\mu \rangle }-\frac{
\langle \mu |(\eta ^{+})^rH\eta ^r|\mu \rangle }{\langle \mu |(\eta
^{+})^r\eta ^r|\mu \rangle }.  \label{gaps}
\end{equation}
For $\mu =0$ we evidently have Yang's functions for which all the gaps are
strictly equal to E. For $\mu \neq 0$ we expect a structure in $\triangle
_r(|\mu |^2)$. In fact all the quantities in (\ref{gaps}) can be calculated
using only ${\cal N}({\vert \mu \vert }^2 )$ and $\langle \mu
|H|\mu \rangle $ (\ref{muhmu}). For a general operator $Q$ we can calculate $
\langle \mu |(\eta ^{+})^rQ\eta ^r|\mu \rangle $ through the relation 
\begin{equation}
\langle \mu |(\eta ^{+})^rQ\eta ^r|\mu \rangle =(1+|\mu |^2)^{-M}\frac{
\partial ^r}{\partial (\mu ^{*})^r}\frac{\partial ^r}{\partial \mu ^r}\left[
(1+|\mu |^2)^M\langle \mu |Q|\mu \rangle \right] ,  \label{Qeval}
\end{equation}
which indicates that for $\rho \equiv |\mu |^{2\text{ }}$ the generating
function for the matrix elements of $Q$ between the depleted states is
proportional to $(1+\rho )^M\langle \mu |Q|\mu \rangle $, which for $Q\equiv
1$ and $Q\equiv H$ furnishes all the input for Eq. (\ref{gaps}).

The detailed analysis of  Eq.(\ref{gaps}) confirms a very interesting structure
of $\triangle _r$ as a function of $\rho $, M and r. The precise description
will be presented elsewhere but we note here  that the gaps as a function of 
$\rho $ go through a maximum for $r\approx $ $\frac M2$ which in turn
disappears for $r>\frac M2$. This confirms that the half-filling point ($
N=\frac M2$) plays a special r\^{o}le for the Hubbard model.
Equation  \ref{Qeval} may be used to obtain the following simple result for the energy dispersion in 
a coherent pairing state:
\begin{equation}
\frac{(\triangle H)^2}{\langle \mu |H|\mu \rangle ^2}=\frac \rho M 
\label{1overM}
\end{equation}
where $\rho = |\mu |^2.$
This indicates that the energy fluctuations are normal in the thermodynamic sense, as  in the 
grand canonical ensemble.  Similarly, in the first depleted state ${\vert} \mu;1 \rangle$
\begin{equation}
\frac{{{(\triangle H)_1}^2}}{\langle H \rangle {_1} ^2}=\frac {\rho (2+2M\rho-M\rho^2 +M^2 
\rho^2)}{(M-1)(1-\rho+M \rho)^2 }.
\label{first}
\end{equation}
Note that the dispersion in the first depleted state Equation(\ref{first}) is always greater than that in 
an SCS, Equation(\ref{1overM}).  Analogous, if more complex, results hold for higher depleted 
states. \\ \\
{\underline{Time evolution of the Coherent Pairing States} 
Since the coherent pairing states are not eigenstates of the Hubbard hamiltonian, they possess a 
non-trivial time dependence. This time evolution is easily calculable via the time-dependent 
Schroedinger equation  due to the simple algebraic structure of the model. For the case of a 
conventional coherent state satisfying $a|z>=z|z>$ evolving under the action of a hamiltonian 
$H=\omega( a^{+} a + \frac{1}{2})$, the evolution is simply expressed by the propagator
($\hbar = 1$)
\begin{equation}
|<z(0)|z(t)>| =exp(|z|^2[\cos{ \omega t} - 1]). 
\label{zt}
\end{equation} 
In the case of the coherent pairing state Eq.(\ref{nscs}), the analogous result, with Eq(\ref{spec}) 
is 
\begin{equation}
|<\mu (0)|\mu (t)>| =|\frac{(1+|\mu |^{2} e^{itE})^{M}}{(1+|\mu |^{2} )^{M}}|. 
\label{mut}
\end{equation} 
In the limit $M\rightarrow \infty$,
$\mu \rightarrow \frac{z}{\sqrt{M}}$  
(corresponding to the group contraction  $\eta \rightarrow \sqrt{M} a^{+}$, compare 
Eq.(\ref{SU2mod}) ) 
we recover the conventional (bosonic) case (\ref{zt}). \\ \\
{\underline{Off-Diagonal Long-Range Order (ODLRO)}  } The presence of Off-Diagonal Long 
Range Order (ODLRO)\cite{Yang62} is detected by the nonvanishing of correlators such as 
$<a_s^ {\dagger}b_s^ {\dagger}b_r a_r>$ as $\vert r-s \vert \rightarrow \infty$.  Yang has shown 
that his states display ODLRO which, in the thermodynamic limit, is proportional to $n_2(1-n_2)$, 
where $n_2 \equiv N/M$ is the pair density.  We may similarly show that our SCS states $\vert 
\mu \rangle$, $\mu \neq 0$, exhibit ODLRO, also proportional to $\overline{n_2}(1-
\overline{n_2})$ where the average pair density $\overline{n_2}=<N>/M$.  Additionally, the states 
$\vert \mu;r \rangle$ exhibit ODLRO and all the results reduce to those of Yang for $\mu =0$ 
\cite{PSM}.  Thus the states  $\vert \mu\rangle$ , $\vert \mu;r \rangle$ are superconducting for all 
$\mu$ and $r$.  It is worth noting that although $\langle \psi_N \vert \eta \vert \psi_N\rangle = 0$ , 
which makes $\eta$ unsuitable for defining  an order parameter in the usual sense, $\langle \mu 
\vert \eta \vert \mu\rangle  \neq 0$ as in the analogous BCS case. \\ \\
{\underline{Relation to Mean Field Theory}  } We may now write  a mean-field version of the 
Hubbard Hamiltonian $H^{MF}=\sum_k
H_k$ 
\begin{equation}
H_k = E_k ({a_k}^{\dagger} a_k +{b_k}^{\dagger} b_k)+ 2W ({\Delta_k}^{*} \eta_k
+\Delta_k {\eta_k}^{\dagger})
\label{hk}
\end{equation} 
with
$\eta_q = \sum_k{a_k b_{q-k}} \; \; \; (q = {\bf \pi})$ and effective energies $E_k$ which include 
the $T_0$ and $T_1$ terms of  Equations(\ref{t0}) and (\ref{t1}).
The spectrum-generating algebra for the Hamiltonian Eqn(\ref{hk}) is $u_k(2)$. 
The non-diagonal terms are the number-non-conserving analogues of a spin-density
wave system (see, for example, \cite{ais1} ) and were already  observed in a multi-phase
$SU(8)$ model\cite{ais2}, where they were called  ``anomalous'' terms, their relation
to the Hubbard model at that time not having been appreciated.  Thus in the
mean-field approximation the dynamical group is $\otimes_k U_k(2)$.  The
associated group parameters (Bogoliubov transformation angles) are $\mu_{{\bf
k}}$, vector analogues of the $\mu$ parameter in the $U(2)$ which is what
remains of the dynamical symmetry in the exact model.\\ \\
{\underline{Extensions beyond the Hubbard Model}  It is possible to demonstrate that the relation 
(\ref{spec}) (with modified
E) can be fulfilled by Hamiltonians other than (1). Yang already mentioned the possibility \cite
{Yang89} of non-local interactions satisfying (\ref{spec}) with
appropriately modified $\eta .$ It is interesting to observe that at least
one case of a truly non-local interaction satisfies (\ref{spec}). It
concerns an extension of the pair-hopping model \cite
{PKM} of the form\cite{PSM} 
\begin{equation}
H_{ph}=T_0+T_1+V\sum_{\langle rs\rangle }\vec{\eta _r}\cdot 
\vec{\eta _s}-\frac V2\sum_rn_r
\end{equation}
which satisfies $\left[ H_{ph}-(T_0+T_1),\eta ^{+}\right] =V\eta ^{+}.$ Still
other extensions are possible for which the coherent pairing state ${\vert}\mu
\rangle $ is a useful tool\cite{PSM}.

\end{document}